# Hydrostatic pressure: A very effective approach to significantly enhance critical current density in granular $Sr_4V_2O_6Fe_2As_2$ superconductor

Babar Shabbir[1], Xiaolin Wang[1*], S. R. Ghorbani[1,2], Shixue Dou[1], Chandra Shekhar[3], and O.N. Srivastava[3]

[1]Spintronic and Electronic Materials Group, Institute for Superconducting & Electronic Materials, Australian Institute for Innovative Materials, University of Wollongong, Wollongong, NSW, 2500, Australia

[2]Department of Physics, Ferdowsi University of Mashhad, Mashhad, Iran

[3]Department of Physics, Banaras Hindu University, Varanasi-221005.

(June 12, 2014)

Pressure is well known to significantly raise the superconducting transition temperature, $T_c$, in both iron pnictides and cuprate based superconductors. Little work has been done, however, on how pressure can affect the flux pinning and critical current density in the Fe-based superconductors. Here, we propose to use hydrostatic pressure to significantly enhance flux pinning and $T_c$ in polycrystalline pnictide bulks. We have chosen $Sr_4V_2O_6Fe_2As_2$ polycrystalline samples as a case study. We demonstrate that the hydrostatic pressure up to 1.2 GPa can not only significantly increase $T_c$ from 15 K (underdoped) to 22 K, but also significantly enhance the irreversibility field, $H_{irr}$, by a factor of 4 at 7 K, as well as the critical current density, $J_c$, by up to 30 times at both low and high fields. It was found that pressure can induce more point defects, which are mainly responsible for the $J_c$ enhancement. In addition, we found that the transformation from surface pinning to point pinning induced by pressure was accompanied by a reduction of anisotropy at high temperatures. Our findings provide an effective method to significantly enhance $T_c$, $J_c$, $H_{irr}$, and the upper critical field, $H_{c2}$, for other families of Fe-based superconductors in the forms of wires/tapes, films, and single crystal and polycrystalline bulks.

Iron based superconductors have revealed wonderful superconducting properties, including high values of critical temperature ($T_c$), critical current density ($J_c$), upper critical field ($H_{c2}$), and irreversibility field ($H_{irr}$). They also exhibit low anisotropy and very strong pinning, which gives rise to high $J_c$ (~$10^6$ A/cm$^2$) in both single crystals and thin films at both low and high fields [1-10]. The $J_c$ and its field dependence in polycrystalline bulks and tapes/wires, however, are still lower than what is required for practical applications. Enhancement of $J_c$ or flux pinning using various approaches has always been a main focus of research with a view to large current and high field applications. So far, three main methods have been used to increase the $J_c$ in cuprates, MgB$_2$, and iron based superconductors: 1) texturing processes to reduce the mismatch angle between adjacent grains and thus overcome the weak-link problem in layer-structured superconductors; 2) high energy ion implantation or irradiation to introduce point defect pinning centres; and 3) introducing point pinning centres by chemical doping. $J_c$ values achieved by the irradiation method have reached as high as $10^6$-$10^7$ A/cm$^2$ for both low and high fields in single crystals and thin films [11-13]. This method is not ideal, however, for $J_c$ enhancement in polycrystalline pnictide superconductors.

As is well known, the weak-link issue is the predominant factor causing low $J_c$, especially at high fields in pnictide polycrystalline samples, which must be overcome. In order to improve the $J_c$ and its field dependence in granular superconductors, the following prerequisites should be met: i) strong grain connectivity; ii) introduction of more point defects inside grains; and iii) $T_c$ enhancement, which can increase the effective superconducting volume as well as $H_{irr}$ and $H_{c2}$.

We have taken into account that the following facts relating to flux pinning mechanisms must be addressed before an effective method is introduced for polycrystalline pnictide superconductors. The coherence length is very short ($\xi \approx$ a few nm), so elimination of weakly linked grain boundaries is important to achieve high $J_c$ [14]. The nature of the pinning mechanism plays a vital role in $J_c$ field dependence. It is noteworthy that a high pinning force can boost pinning strength and, in turn, leads to higher values of $J_c$. The ideal size of defects for pinning should be comparable to the coherence length [15]. Therefore, point defect pinning is more favourable than surface pinning, as its pinning force is larger than for surface pinning at high field, according to the Dew-Hughes model [16]. Therefore, it is very desirable to induce more point defects in superconductors. Although chemical doping and high energy particle irradiation can effectively induce point defects and enhance $J_c$ in high fields, $T_c$ and low field $J_c$ deteriorate greatly for various types of superconductors. Therefore, the ideal approach should be the one which can induce more point defects, with increased (or at least at no cost of) superconducting volume and $T_c$, as well as strongly linked grain boundaries.

Hydrostatic pressure has been revealed to have a positive effect on $T_c$ in cuprate and pnictide superconductors. For instance, high pressure of 150 kbars can raise $T_c$ of Hg-1223 significantly from 135 K to a record high 153 K [17]. The $T_c$ of hole doped (NdCeSr)CuO$_4$ was increased from 24 to 33 K at 3 GPa by changing the apical Cu-O distance [18]. The enhancement of $T_c$ for YBCO is more than 10 K at 2 GPa [19]. Excitingly, pressure also shows positive effects on $T_c$ for various pnictide superconductors. Pressure can result in improvement of $T_c$ from 28 to 43 K at 4 GPa for LaOFFeAs [20]. For Co doped NaFeAs, the maximum $T_c$ can reach as high as 31 K from 16 K at 2.5 GPa [21]. Pressure can also enhance the $T_c$ of La doped Ba-122 epitaxial films up to 30.3 K from 22.5 K, due to the reduction of electron scattering and increased carrier density caused by lattice shrinkage [22]. A huge enhancement of $T_c$ from 13 to 27 K at 1.48 GPa was observed for FeSe, and it reached the high value of 37 K at 7 GPa [23, 24].

Beside the above-mentioned significant pressure effects on $T_c$ enhancement, pressure can have more advantages that are relevant to the flux pinning compared to other methods. 1) It always reduces the lattice parameters and causes the shrinkage of unit cells, giving rise to the reduction of anisotropy. 2) Grain connectivity improvement should also be expected, as pressure can compress both grains and grain boundaries. 3) The existence or formation of point defects can be more favourable under pressure, since it is well known that the formation energy of point defects decreases with increasing pressure. 4) Pressure can cause low-angle grain boundaries to migrate in polycrystalline bulk samples, resulting in the emergence of giant grains, sacrificing surface pinning thereafter. Hence, a higher ratio of point pinning centres to surface pinning centres is expected due to the formation energy and migration of grain boundaries under pressure. 5) The significant enhancement of $T_c$, as above-mentioned, means that superconducting volumes should be increased greatly below or above the $T_c$ without pressure. Moreover, the $H_{c2}$, $H_{irr}$, and $J_c$ have to be enhanced along with the $T_c$ enhancement. These are the motivations of our present study on the pressure effects on flux pinning and $J_c$ enhancement in polycrystalline pnictide bulks. We anticipated that hydrostatic pressure would increase the superconducting volume, $H_{irr}$, and $H_{c2}$ due to $T_c$ enhancement, increase the point defects, improve grain connectivity, and reduce the anisotropy in pnictide polycrystalline bulk samples.

There is some evidence for $J_c$ enhancement under pressure in YBa$_2$Cu$_3$O$_{7-x}$ (YBCO) single crystal, which emphasizes the pressure effects on transport $J_c$ for different angle grain boundaries. A recent report also shows enhanced $J_c$ in a pnictide single crystal which is free of grain boundaries [25-27]. As mentioned earlier, polycrystalline superconducting materials are commonly used in practical applications, as they are easy to fabricate at low cost as compared to single crystals/thin films. Their superconducting performance is hindered by grain boundaries, however, due to granularity. Therefore, it is more important to use an efficient approach to enhance the $J_c$ in polycrystalline bulk samples. In this study, we chose a polycrystalline Sr$_4$V$_2$O$_6$Fe$_2$As$_2$ sample to demonstrate the significant effects of the hydrostatic pressure on flux pinning and the significant enhancement of $J_c$ and $T_c$ in this granular sample. It has been reported that the $T_c$ for this compound can range from 15 - 30 K, depending on fabrication process and carrier concentration [28]. Generally, superconductors under pressure have dome-like plots for $T_c$ vs. pressure, so we chose a Sr$_4$V$_2$O$_6$Fe$_2$As$_2$ sample with the low $T_c$ of 15 K for the proposed pressure effect investigation to ensure a clear pressure effect on $T_c$ [29]. Our results show that pressure can enhance the $J_c$ by more than 30 times at 6 K and high fields in polycrystalline Sr$_4$V$_2$O$_6$Fe$_2$As$_2$, along with $T_c$ enhancement from 15 to 22 K at 1.2 GPa and $H_{irr}$ enhancement by a factor of 4. Our analysis shows that pressure induced point defects inside the grains are mainly responsible for the flux pinning enhancement.

Figure 1 shows the temperature dependence of the zero-field-cooled (ZFC) and field-cooled (FC) moments for Sr$_4$V$_2$O$_6$Fe$_2$As$_2$ at different pressures. Pressure causes little change to the field-cooled branch, indicating that strong pinning is retained under pressure. The $T_c$ without pressure is about 15 K, very similar to that of underdoped samples reported for Sr$_4$V$_2$O$_6$Fe$_2$As$_2$ bulks [28]. Pressure enhances $T_c$ linearly from 15.3 K for $P$ = 0 GPa to 22 K for $P$ = 1.2 GPa, with the pressure coefficient, $dT_c/dP$ = 5.34 K/GPa.

The $M$-$H$ curves measured under different pressures indicate that the moment increases with increasing pressure. The field dependence of $J_c$ at different temperatures obtained from the $M$-$H$ curves by using Bean's model under different pressures is shown in a double-logarithmic plot [i.e. Figure 2(a)]. The remarkable effect of pressure towards the enhancement of $J_c$ can be clearly seen. For $P$ = 1.2 GPa, the $J_c$ is significantly enhanced by more than one order of magnitude at high fields at 4 and 6 K, respectively, as shown in Fig. 2(b).

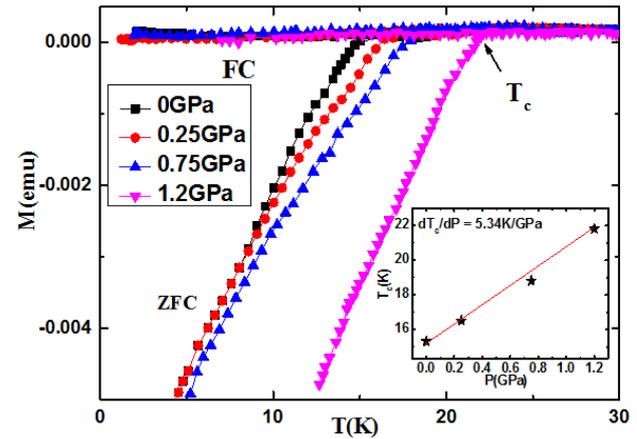

Figure 1: Temperature dependence of ZFC and FC moments at different pressures for Sr$_4$V$_2$O$_6$Fe$_2$As$_2$. The inset shows the pressure dependence of $T_c$.

The $J_c$ at 6 K as a function of pressure at different fields is plotted in Fig. 3. The solid lines in Fig. 3 show linear fits to the data, which give the slopes (i.e. $d(\ln J_c)/dP$) of 1.09, 1.69, and 2.30 GPa$^{-1}$ at 0, 2, and 4 T, respectively, indicating that the effects of pressure towards the enhancement of the $J_c$ are more significant at high fields.

We also found that the $H_{irr}$ of Sr$_4$V$_2$O$_6$Fe$_2$As$_2$ is greatly increased by pressure. As shown in Fig. 4, the $H_{irr}$ increases gradually with pressure and rises to 13 T from 3.5 T at 7 K. The $J_c$ vs. reduced temperature (1-$T/T_c$) at zero field and different pressures is plotted in Figure 5, which shows a rough scaling behaviour as $J_c \propto (1 - T/T_c)^\beta$ at different pressures. The slope of the fitting line, $\beta$, depends on the magnetic field. The exponent $\beta$ (i.e. slope of the fitting line) is found to be 2.54, 2.73, 2.96, and 3.13 at 0, 0.25, 0.75, and 1.2 GPa, respectively. According to Ginzburg-Landau theory, the exponent "$\beta$" is used to identify different vortex pinning mechanisms at specific magnetic fields. It was found that $\beta$ = 1 for non-interacting vortices, while $\beta \geq 1.5$ indicates the core pinning mechanism [30]. The different values of β (i.e. 1.7, 2, and 2.5) were also reported for YBCO films which show the functioning of different core pinning mechanisms [31, 32]. In addition, the exponent $\beta$ values that we obtained are higher at higher pressures in our sample, For polycrystalline samples, high pressure can modify the grain boundaries through reducing the tunnelling barrier width and changing the tunnelling barrier height. The Wentzel-Kramers-Brillouin (WKB) approximation applied to a potential barrier gives the following simple expressions [33] :

$$J_c = J_{co} \exp(-2kW) \quad (1)$$

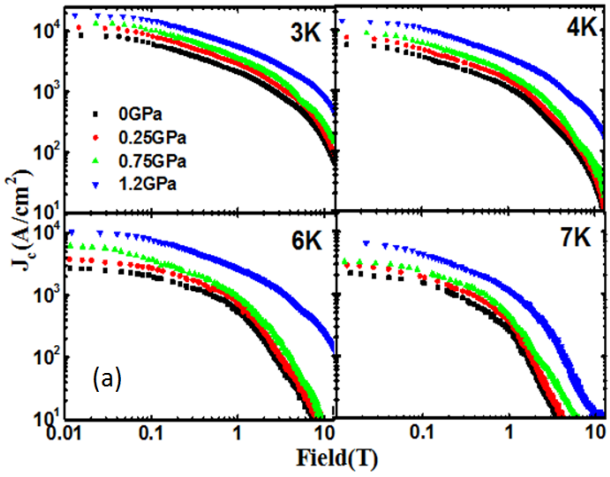

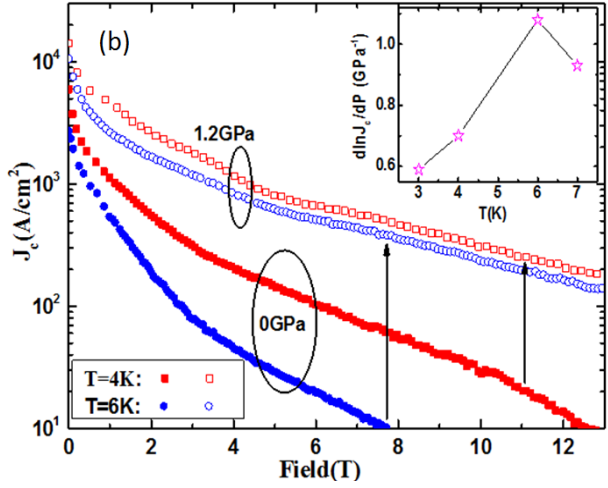

Figure 2(a): Field dependence of $J_c$ under different pressures at 3, 4, 6, and 7 K. (b). Comparison of $J_c$ at 0 and 1.2 GPa at 4 and 6 K. The inset shows $d(\ln J_c)/dP$ versus temperature, indicating enhancement of $J_c$ at a rate of 1.08 $GPa^{-1}$ at zero field.

Where $W$ is the barrier width, $k = (2mL)^{1/2}/\hbar$ is the decay constant, which depends on the barrier height $L$, $\hbar$ is the Planck constant, and $J_{c0}$ is the critical current density for samples with no grain boundaries. The relative pressure dependence of $J_c$ can be obtained from Eq. (1) as:

$$\frac{d\ln J_c}{dP} = \frac{d\ln J_{c0}}{dP} - \left[\left(\frac{d\ln W}{dP}\right)\ln\left(\frac{J_{c0}}{J_c}\right)\right] - \frac{1}{2}\left[\left(\frac{d\ln L}{dP}\right)\ln\left(\frac{J_{c0}}{J_c}\right)\right]$$
$$= \frac{d\ln J_{c0}}{dP} + \kappa_{GB}\ln\left(\frac{J_{c0}}{J_c}\right) + \frac{1}{2}\kappa_L\ln\left(\frac{J_{c0}}{J_c}\right) \quad (2)$$

Where the compressibility in the width and height of the grain boundary are defined by $\kappa_{GB} = -d\ln W/dP$ and $\kappa_L = -d\ln L/dP$, respectively. To estimate their contributions to the second and the third terms of Eq. (2) for $J_c$ enhancement, we assume to a first approximation that $\kappa_{GB}$ and $\kappa_L$ are roughly comparable to the average linear compressibility values $\kappa_a = -d\ln a/dP$ ($\kappa_{GB} \approx \kappa_a$) and $\kappa_c = -d\ln c/dP$ ($\kappa_L \approx \kappa_c$) of $Sr_4V_2O_6Fe_2As_2$ in the FeSe plane, where $a$ and $c$ are the in-plane and out-of-plane lattice parameters, respectively. We assume to a first approximation that $\kappa_a = -d\ln a/dP = -0.029$ $GPa^{-1}$, $\kappa_c = -d\ln c/dP = -0.065$ $GPa^{-1}$ [24], and the in-plane $J_{c0} \cong 10^5$ $A/cm^2$ for a sample with no grain boundaries (single crystal) [34]. Setting $J_c \cong 2 \times 10^3$ $A/cm^2$ at the temperature of 6 K and ambient pressure, we find that $(-d\ln W/dP)\ln(J_{c0}/J_c) \approx 0.11$ and $-0.5(d\ln L/dP)\ln(J_{c0}/J_c)] \approx 0.13$, indicating stronger improvement of $J_c$ with temperature at high pressures.

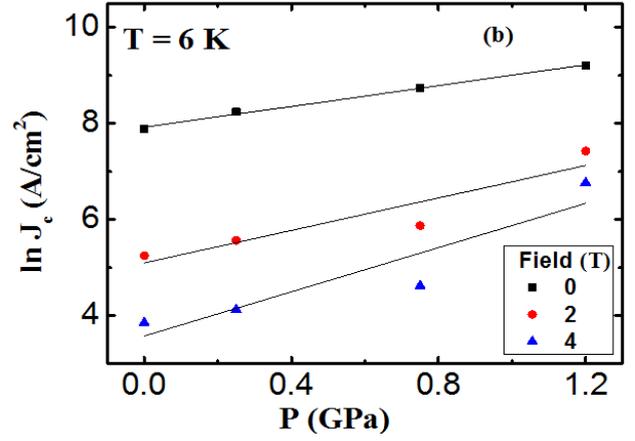

Figure 3: Pressure dependence of critical current density (logarithmic scale) at 0, 2, and 4 T at the temperature of 6 K..

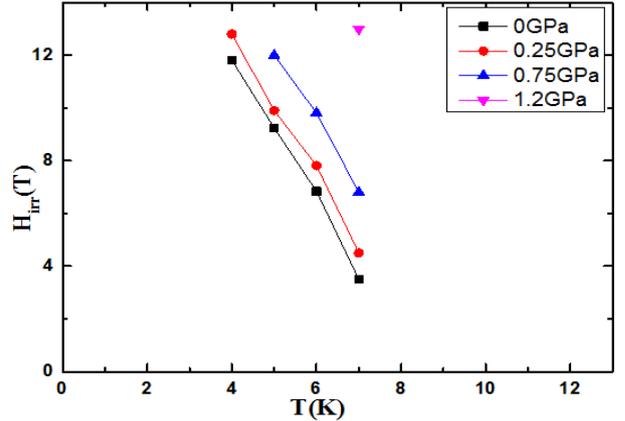

Figure 4: $H_{irr}$ vs. $T$ for different pressures.

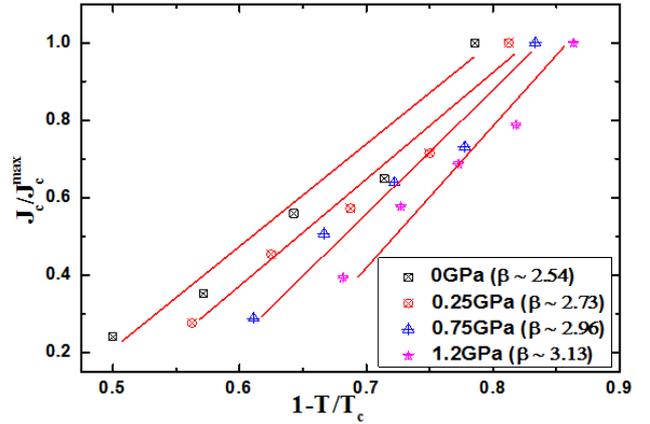

Figure 5: Logarithmic plot of $J_c$ as a function of reduced temperature at different pressures and fields.

with both values adding up to 75% less than the above experimental value $d\ln J_c/dP = 1.08$ $GPa^{-1}$. This result suggests that the origin of the significant increase in $J_c(T)$ under pressure does not arise from the compression of the grain boundaries. Therefore, Eq. (2) suggests that the main reason for the rapid increase of $J_c$ with pressure is through point defects induced under pressure, i.e., $d\ln J_{c0}/dP$ is responsible for approximately 75% of the total increase in the $J_c$ with pressure.

In order to further understand the $J_c$ enhancement under pressure, as shown in Fig. 3(a), the pinning force $F_p = B \times J_c$ is calculated, and the scaling behaviour for the normalized pinning force $f_p = F_p/F_{p,max}$, is analysed for $h = H/H_{irr}$.

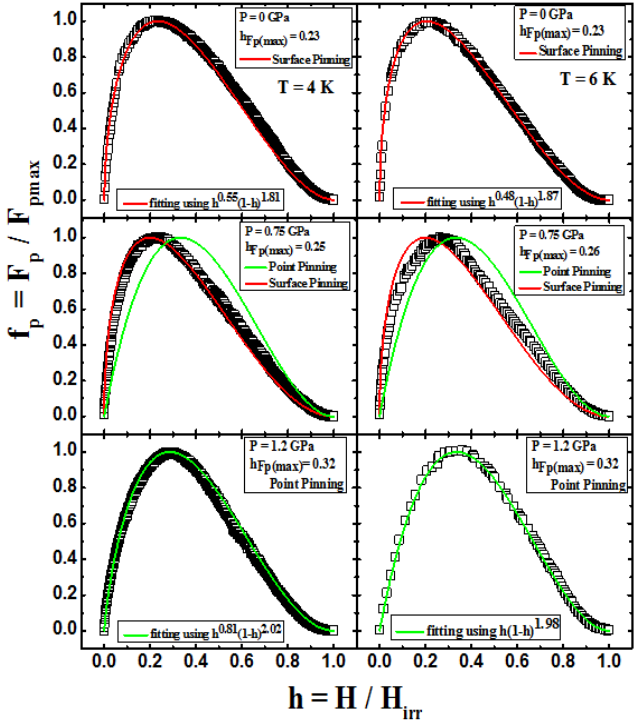

Figure 6: Plots of $f_p$ vs. $H/H_{irr}$ at different pressures (0, 0.75, and 1.2 GPa) for 4 (left) and 6 K (right) temperature curves. The experimental data is fitted through the Dew-Hughes model, and the parameters are shown.

The results are shown in Fig. 6 at 4 and 6 K under 0, 0.75, and 1.2 GPa. For the scaling, we can use the Dew-Hughes formula, i.e. $f_P(h) = A h^p (1-h)^q$, where $p$ and $q$ are parameters describing the pinning mechanism [16]. In this model, $p = 1/2$ and $q = 2$ describes surface pinning while $p = 1$ and $q = 2$ describes point pinning, as was predicted by Kramer [35]. At ambient pressure in the temperature range of 3-7 K, the best fits of the curves are obtained with $p = 0.51 \pm 0.03$, $q = 1.86 \pm 0.03$, which suggests that surface pinning is the dominant pinning mechanism in our sample. At 1.2 GPa, the best obtained values for $p$ and $q$ were $0.9 \pm 0.1$ and $2 \pm 0.1$, respectively, within the studied temperature range. This means that the dominant pinning mechanism is normal core point pinning for high pressures. Therefore, our results show that the pressure has induced a clear transformation from surface to point pinning.

Moreover, it is noteworthy that pressure can induce a reduction in anisotropy. At high temperatures, the pressure dependence of $T_c$, unit cell volume ($V$), and anisotropy ($\gamma$) are interconnected through the following relation [36];

$$\frac{-\Delta T_c}{T_c} = \frac{\Delta V(T_c)}{V(T_c)} + F(\gamma) \quad (3)$$

Where $F(\gamma) = \frac{\Delta f(\gamma(T_c))}{f(\gamma(T_c))}$ and $f(\gamma(T_c))$ is a function of the anisotropy. We can use the bulk modulus ($K \approx 62$ GPa) of a similar superconductor, i.e., $SrFe_2As_2$, to estimate $\Delta V(T_c)/V(T_c)$, which is found to be $\sim -0.016$ at $\Delta P = 1$ GPa, as it can be related to the bulk modulus as $\Delta V/V = -\Delta P/K$ [37]. Our experimental results yield a value of 0.486 for $\Delta T_c/T_c = [T_c(P) - T_c(0)]/T_c(0)$. By using these results, we can obtain $F(\gamma) \approx -0.47$. Thus, we can conclude that the anisotropy has been reduced by almost half at high temperature by pressure. The decrease in the unit cell parameters suppresses its volume, leading to an increase in the Fermi vector $k_F = (3\pi^2 N/V)^{1/3}$, where $N$ is the total number of electrons in the system. The increase in the Fermi vector promotes enhancement of the coherence length along the $c$-axis ($\xi_c = \hbar^2 K_F/\pi m \Delta$ where $\Delta$ is the uniform energy gap) which, in turn, leads to the suppression of anisotropy.

In summary, hydrostatic pressure is a very effective means to significantly enhance $T_c$, $J_c$, $H_{irr}$, and flux pinning in the granular pnictide superconductor $Sr_4V_2O_6Fe_2As_2$. We demonstrate that the hydrostatic pressure can significantly increase $T_c$ from 15 to 22 K, as well as increasing $J_c$ by up to 30 times at both low and high field and increasing $H_{irr}$ by a factor of 4 at P=1.2 GPa. Pressure introduces more point defects inside grains, so that it is mainly responsible for $J_c$ enhancement. Our findings provide an effective method to significantly enhance $T_c$, $J_c$, $H_{irr}$, and $H_{c2}$ for other families of Fe-based superconductors in the forms of wires/tapes, films, and single and polycrystalline bulks.

X.L.W. acknowledges the support from the Australian Research Council (ARC) through an ARC Discovery Project (DP130102956) and an ARC Professorial Future Fellowship project (FT130100778).

*Correspondence should be addressed to Xiaolin Wang
Email: xiaolin@uow.edu.au